\documentclass[useAMS,usenatbib]{mn2e}
\usepackage{graphicx}

\title[The first release of data from the {\it Herschel} ATLAS: the SPIRE images]{The first release of data from the {\it Herschel} ATLAS: the SPIRE\thanks{{\it Herschel} is an ESA space observatory with science instruments provided by European-led Principal Investigator consortia and with important participation from NASA} images}
\author[E.~Pascale et al.]{
E.~Pascale,$^{1}$\thanks{E-mail: enzo.pascale@astro.cf.ac.uk}
R.~Auld,$^{1}$
A.~Dariush,$^{1}$
L.~Dunne,$^{2}$
S.~Eales,$^{1}$
S.~Maddox,$^{2}$
P.~Panuzzo,$^{3}$
\and
M.~Pohlen,$^{1}$
D.~J.~B.~Smith,$^{2}$ 
S.~Buttiglione,$^{4}$
A.~Cava,$^{5}$
D.~L.~Clements,$^{6}$
A.~Cooray,$^{7}$
\and
S.~Dye,$^{1}$
G.~de~Zotti,$^{4,\,8}$
J.~Fritz,$^{9}$
R.~Hopwood,$^{10}$
E.~Ibar,$^{11}$
R.~J.~Ivison,$^{11,\,12}$
\and
M.~J.~Jarvis,$^{13}$
L.~Leeuw,$^{14}$
M.~L\'opez-Caniego,$^{15}$
E.~Rigby,$^{2}$
G.~Rodighiero,$^{16}$
D.~Scott,$^{17}$
\and
M.~W.~L.~Smith,$^{1}$ 
P.~Temi$^{18}$
M.~Vaccari,$^{4}$
and I.~Valtchanov$^{19}$
\vspace*{5mm} \\
$^{1}$School of Physics and Astronomy, Cardiff University, Queens Buildings, The Parade, Cardiff CF24 3AA, UK \\
$^{2}$School of Physics and Astronomy, University of Nottingham, Nottingham NG7 2RD, UK\\
$^{3}$CEA, Laboratoire AIM, Irfu/SAp, F-91191 Gif-sur-Yvette, France \\
$^{4}$INAF-Osservatorio Astronomico di Padova, Vicolo dell'Osservatorio 5, I-35122 Padova, Italy \\
$^{5}$Instituto de Astrof\'isica de Canarias, Calle V\'ia L\'actea, E38205, La Laguna, Espa\~na\\
$^{6}$Astrophysics Group, Imperial College, Blackett Laboratory, Prince Consort Road, London SW7 2AZ, UK\\
$^{7}$Department of Physics and Astronomy, University of California, Irvine, CA 92697, USA\\
$^{8}$SISSA, Via Bonomea 265, I-34136 Trieste, Italy\\
$^{9}$Sterrenkundig Observatorium, Universiteit Gent, Krijgslaan 281 S9, B-9000 Gent, Belgium \\
$^{10}$Deptartment of Physics and Astronomy, The Open University, Milton Keynes MK7 6AA, UK\\
$^{11}$UK Astronomy Technology Centre, Royal Observatory, Blackford Hill, Edinburgh EH9 3HJ, UK\\
$^{12}$Institute for Astronomy, University of Edinburgh, Royal Observatory, Edinburgh EH9 3HJ, UK \\
$^{13}$Centre for Astrophysics Research, STRI, University of Hertfordshire, Hatfield AL10 9AB, UK\\
$^{14}$Astrophysics Branch, NASA Ames Research Center, MS 245-6, Moffett Field, CA 94035, USA \\
$^{15}$Instituto de F\'isica de Cantabria (CSIC-UC), Santander 39005, Spain\\ 
$^{16}$Department of Astronomy, University of Padova, Vicolo Osservatorio 3, I-35122, Padova, Italy \\
$^{17}$Department of Physics \& Astronomy, University of British Columbia, 6224 Agricultural Road, Vancouver, BC V6T~1Z1, Canada\\
$^{18}$ Astrophysics Branch, NASA Ames Research Center, Mail Stop 245-6, Moffett Field, CA 94035, USA \\
$^{19}$Herschel Science Centre, European Space Astronomy Centre, Villanueva de la Ca\~nada, 28691 Madrid, Spain\\
}

\begin{document}
\label{firstpage}



\maketitle

\begin{abstract}
We have reduced the data taken with the Spectral and Photometric Imaging Receiver (SPIRE) photometer on board the {\it Herschel} Space Observatory in the Science Demonstration Phase (SDP) of the {\it Herschel} Astrophysical Terahertz Large Area Survey (H-ATLAS). 
We describe the data reduction, which poses specific challenges, both because of the sheer size of the data, and because only two scans are made for each region. We implement effective solutions to process the bolometric timelines into maps, and show that correlations among detectors are negligible, and that the photometer is stable on time scales up to 250\,s. This is longer than the time the telescope takes to cross the observed sky region, and it allows us to use naive binning methods for an optimal reconstruction of the sky emission. The maps have equal contribution of confusion and white instrumental noise, and the latter is estimated to 5.3, 6.4, and 6.7 mJy\,beam$^{-1}$ (1-$\sigma$), at 250, 350, and 500\,$\umu$m, respectively. 
This pipeline is used to reduce other H-ATLAS observations, as they became available, 
and we discuss how it can be used with the optimal map maker implemented in the {\it Herschel} Interactive Processing Environment (HIPE), to improve computational efficiency and stability. The SDP dataset is available from http://www.h-atlas.org/.
\end{abstract}


\section{Introduction}
The {\it Herschel} Astrophysical Terahertz Large Area Survey \citep[H-ATLAS,][]{eales2010} 
is currently the largest area open-time key project that is being carried out by the {\it Herschel} Space Observatory 
\citep{pilbratt2010}. The 600\,hours of observations will map $\sim 550$\,deg$^2$ of the sky in the 5 far-IR and submillimetric bands of the Photodetector Array Camera \& Spectrometer (PACS) \citep{poglitsch2010} and SPIRE \citep{griffin2010} photometers. The survey of unprecedented sky and spectral coverage, and angular resolution, will target sky regions selected in the direction of the North and South Galactic Poles, and of the Galaxy And Mass Assembly Survey, the GAMA fields \citep{driver2009}.

The first observations of the project have been completed on a 16\,deg$^2$ GAMA target, as part of the 
 initial Science Demonstration Phase (SDP). Here we present the maps obtained with the SPIRE photometer,  and the reduction pipeline, which differs in significant aspects from the standard pipeline described by \cite{griffin2008,griffin2010}.  The SDP data are publicly available\footnote{The SDP H-ATLAS data can be accessed from http://www.h-atlas.org.}, and comprise a set of three SPIRE photometer maps at 250, 350, and 500\,$\umu$m (discussed in this work), and two PACS photometer maps at 100 and 160\,$\umu$m \citep{ibar2010}. Each map has its corresponding error and coverage maps as separate extensions. Sources (point-like and extended) are detected in these maps by \citet{rigby2010}, and used by \citet{smith2010} to identify counterparts in catalogues of UV and near-IR selected sources. 

SPIRE uses arrays of sparsely distributed bolometric detectors to sample the sky radiation through scanning at a specific angle with respect to the bolometer array axes, and the challenge of map making or sky reconstruction is a linear algebra problem which has been extensively investigated in the past \citep[see Section 8 of][for a list of references]{cantalupo2010}. The bolometric data are combined to provide optimal estimates of the sky emission, for which detailed knowledge of the detector noise properties is required. Most experiments (including the High Frequency Instrument on {\sl Planck})
have only a limited number of detectors, sparsely populating their focal planes, 
 but the large format bolometric cameras, which are now available, pose new challenges, as the amount of data is increasing, with potential for correlation among detectors, complicating the analysis \citep{patanchon2008}.
This is of relevance for SPIRE, and the detector noise properties need to be investigated in order to optimally reduce the data into maps.

The choice of a map making technique is dictated by the noise properties of the data. In this paper we analyse the noise in the bolometric timelines, and we use this information to generate two sets of maps: one optimised to study emission from point sources, and the other to study the large-scale structure accessible in the observed sky region. The former have been used in the scientific analyses of the SDP H-ATLAS data published in the 
Herschel First Results special A\&A edition (2010)

This analysis procedure provides a first investigation into the properties of this data-set, and our results can be used in future studies to select appropriate map making techniques, which may improve on
the one used here. However, the pipeline we discuss here will  provide the necessary input to these techniques, in order to optimise their performance.

\section{The SPIRE photometer}
The SPIRE photometer images  the sky in three submillimetric bands centred at 
250, 350, and 500\,$\umu$m. Each band is 30 per cent wide (in $\Delta\lambda/\lambda$) and the 
focal plane is populated with arrays of silicon nitride micromesh ``spider web" bolometers \citep{bock1996, bock1998} operated at a cryogenic temperature of $\la 300$\,mK.
The detectors are coupled to the spacecraft 3.5-m mirror telescope by 
smooth walled feed horns, and 4.5\,K re-imaging optics, which also provide shielding from  stray radiation. The linear distance of adjacent feed horns is $2f\lambda$ \citep{rownd2003} where $f$ is the focal ratio and $\lambda$ is the wavelength.
Compared to its balloon-borne predecessor,
BLAST \citep[][]{pascale2008,devlin2009}, SPIRE 
has better angular resolution, with point spread functions (PSFs) of 
18.1, 25.2, and 36.6\arcsec\ full width at half maximum (FWHM) at 
250, 350 and 500\,$\umu$m, respectively \citep[][]{griffin2010}.
Each array has respectively 139, 88, and 43 light sensitive bolometers, as well as 2 dark bolometers and one resistor. Thermistors on each array monitor temperature variations in the substrate, due to fluctuations in the thermal bath.
The bolometers are AC-biased, and the voltage RMS measured across each detector is proportional to the radiation detected from the sky. The constant of proportionality is the detector's 
responsivity, which is measured by \citet{swinyard2010} from observations of sources with known spectral energy distributions.

\section{The data}
The SDP H-ATLAS data was obtained in parallel mode, with the two Herschel instruments, PACS and SPIRE, operating simultaneously. The H-ATLAS data set consists of two observations - one at the nominal SPIRE scan angle and the other with scans at nearly orthogonal direction ($\sim 85^\circ$), over the same 16\,deg$^2$ square region centered at coordinates RA~=~9$^{\rm h}$~5$^{\rm m}$~30$^{\rm s}$, Dec~=~0$^\circ$~5\arcmin~0\arcsec\@.  Each observation is carried in $\sim 8$\,h by scanning the telescope at the constant angular velocity of 60 arcsec\,s$^{-1}$ over great circles. The scan lines are separated by 155 arcsec in order to achieve good coverage for large fields for both SPIRE and PACS maps. During the scanning the SPIRE bolometers are sampled\footnote{For more details please consult SPIRE and PACS-SPIRE Parallel Mode Observers' Manuals available at the Herschel Science Centre web pages: http://herschel.esac.esa.int.} at 10\,Hz.


\section{Map making method}\label{sec:mapmaking}
Map making is the operation which provides one estimate of the sky emission at the observed wavelength, given the data, and their noise properties.
The data consist of bolometer timelines:
\begin{equation}
  d_i(t) = A_{ip}(t)\,S_p + n_i(t)
\end{equation}
where $t$ is the sample's time, $i$ is the detector index in the 250, 350, or 500\,$\mu$m array, $n_i(t)$ is its noise, $S_p$ is the sky emission from the map pixel $p$, and $A_{ip}(t)$ is the pointing 
matrix, the elements of which give the weight of the contribution of pixel $p$ to the detector $i$ at time $t$. Summation over repeated indices is assumed. Throughout this discussion it is assumed that $S_p$ is 
the convolution of the sky emission with the telescope PSF.

The well known Generalised Least Squares (GLS) estimator of the sky \citep[e.g.][]{lupton1993} is
\begin{equation}
 \tilde{S} =  (A^t N^{-1} A)^{-1} AN^{-1}d \label{eq:mmml}
\end{equation}
where all indices have been dropped for clarity, and $N$ is the noise covariance matrix of the data. It 
is computationally challenging to apply this equation, especially when there are correlations among detectors, and a number of map making 
techniques have been considered in the literature to deal with this problem. However, when the noise is white and uncorrelated among detectors, this equation reduces to a simple, unweighted binning
(naive binning), because $N$ becomes a scalar matrix and its eigenvalue is the timeline variance.

When the timeline noise has a $1/f$-type component, high-pass filters can be used to whiten, or suppress a drifting baseline. 
This results in an attenuation of the map at angular scales larger than $\lambda_{\rm c} = f_{\rm c}/\omega$, where $f_{\rm c}$ is the filter's cutoff frequency, and $\omega$ the scan velocity. 
As shown later in Section~\ref{sec:noise_analysis}, the outstanding stability of the SPIRE photometer makes it possible to set $\lambda_{\rm c}$ at the map scale. 
Therefore essentially all Fourier modes are preserved in the telescope scan direction. Cross-scan modes, however, are suppressed, and can only be recovered by scanning at an orthogonal angle. This can be seen when studying 
Equation~\ref{eq:mmml} in Fourier space, as discussed in \citet{patanchon2008}. 
The two SDP H-ATLAS  observations can be reduced into a map using a 
naive binning, and simple co-addition. Since each map preserves only one set of modes, the final, coadded map suppresses (by up to a factor 2) the large scales, and therefore it is no longer equivalent to the GLS solution of Equation~\ref{eq:mmml}. It is, however, possible to recover the suppressed modes, since all the necessary information is present and the filtering effect can be deconvolved. This is demonstrated in Section \ref{sec:sky_reconstruction}, by using the map maker transfer function.

\section{Data preprocessing}\label{sec:datapreprocessing}
\begin{figure*}
\includegraphics[angle=-90,width=0.8\linewidth]{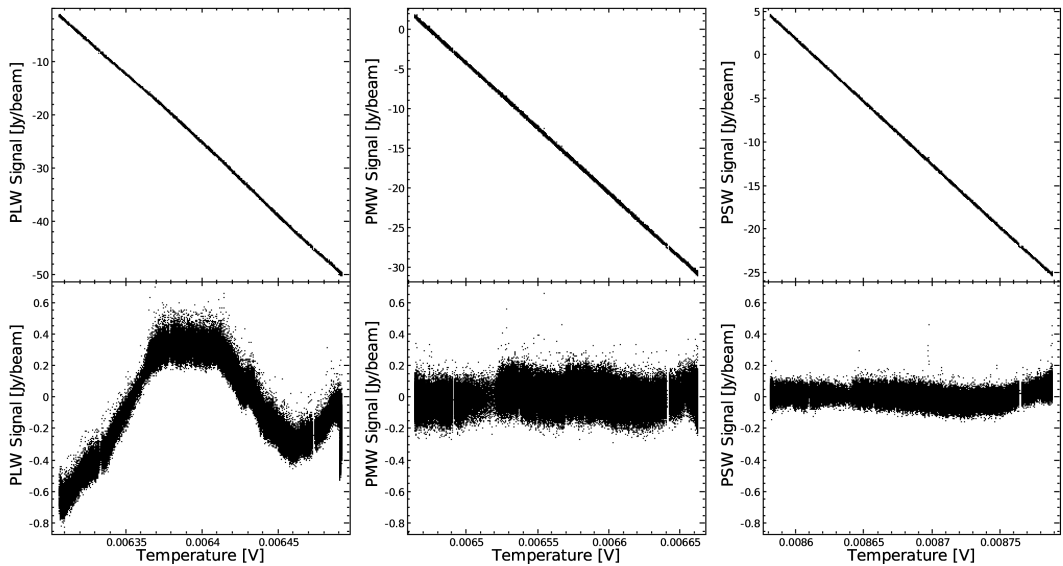}

\caption{The tight correlation between the light SPIRE detectors and one of the
temperature sensors of each array is shown in scatter plots of bolometer timelines
vs. temperature in the three top panels. 
The  bottom panels show the same scatter plot, after the temperature 
correlation is removed from the timelines as discussed in Section \ref{sec:correlatednoise}. 
Detectors in the 250 and 350\,$\mu$m arrays 
are stable. The detectors in the 500\,$\mu$m array are less so, and show a residual 
slow drift which is absent in the dark channel (see Section \ref{sec:correlatednoise} for a discussion). The small gaps in the data
correspond to temporal gaps and discontinuities in the timelines.}
\label{fig:corr}
\end{figure*}
Each detector timeline is preprocessed using a customised version of the SPIRE 
standard pipeline \citep{griffin2008, griffin2010}, implemented in the {\it Herschel} Interactive Processing Environment (HIPE), which is run in an interactive mode, starting from
the Level-0.5 data products (DP). These are a set of FITS binary tables  grouping together the uncalibrated light and dark detector timelines, and all relevant housekeeping information. Each table contains data from just one scan, so there is a large number of DP tables for each observation. 

 Pointing information is associated with each bolometric sample using the standard pipeline, and the timelines are converted to flux density per beam by applying the calibration of \citet{swinyard2010} (the calibration is further discussed in Section \ref{sec:calibration}).
 
Glitches (mostly cosmic ray events, or Bremsstrahlung within the instrument) in the bolometer and thermistor data are flagged, and replaced with a constrained realization of white noise. Two alternative deglitchers are used (alternative in the sense that they are available, but not used by default in the standard pipeline) to identify these transients. One localizes synchronous transients among detectors in a given array, and the other is a sigma-kappa deglitcher which looks for isolated events. This allows the detection of a larger number of low signal-to-noise glitches which could create false point-sources in the final maps, if not corrected. 

DC-offset jumps of unknown origin are occasionally present in the thermistor and bolometer timelines. These jumps cause temporary de-correlation of the thermistor and bolometer timelines, so that usual attempts to correct for the temperature drift in the bolometer thermal bath (see the next section) fail. Unless corrected for, these jumps cause large scale-artifacts (stripes) in the final maps, at several sigma level.

At the time of these observations, the standard pipeline did not have the capability to search for and correct the DC-offset jumps. Therefore transient events have been identified by eye and compensated for by shifting the affected timelines by an appropriate DC-offset, which is estimated from the data.

The electronic and bolometric instrumental transfer functions are deconvolved from the timelines (using the standard pipeline) to recover a flat spectral response, and to remove filtering phase delays.

Data products of each scan are linked together to provide continuous timelines for each observation (including spacecraft turnaround). There is a clear advantage in doing this in the analysis which follows, because signal processing Fourier methods (filtering, but also power spectral analyses) perform better, and the unavoidable artifacts in the data are confined at the map edges, corresponding to the beginning and the end of each observation. Temporal gaps in the data acquisition, corresponding to the SPIRE PCAL source flashes and other events, are filled with a constrained realization of white noise, and flagged accordingly. Flagged data are not projected into maps, but filling the gaps with noise improves the stability of filters, and reduces pollution of good data at the edges of gaps.

The final preprocessed timeline is a clean (without glitches or DC jumps) and calibrated data-set, with associated pointing information, which can be binned into a map.

\section{Correlated noise}\label{sec:correlatednoise}
Each light detector timeline contains signal from the sky, together with noise. In a large format array the noise has two components. One is uncorrelated among detectors, the other arises from common signals:
 \begin{equation}
  d_i(t) = A_{ip}(t)\,S_p + n_i(t) + \alpha_i C(t)
\end{equation}
where $C(t)$ is a signal common among detectors,  $\alpha_i$ is a detector-specific constant, and $n_i(t)$ is the uncorrelated (among detectors) noise, which can have a (time--correlated) $1/f$-type spectrum. Here the assumption of only one common signal is made, even though in general there could be several components with different amplitudes. It is not possible to reconstruct sky angular modes with low signal-to-noise corresponding to frequencies dominated by $1/f$-type noise, because coloured noise does not integrate down with time. Since common signals have typically a $1/f$-type power spectrum, they need to be estimated and subtracted at all scales of interest \citep{patanchon2008, pascale2009}.

The dominant common signal in the SPIRE arrays correlates well with the substrate temperature measured by the array thermistors, as shown in the top panels of Fig.~\ref{fig:corr}, where representative detector timelines in the 250, 350, and 500\,$\mu$m arrays are plotted against the corresponding thermistor timelines.
The Pearson correlation coefficients between bolometers and thermistors are all greater than 99 per cent.
The coefficients $\alpha_i$ are estimated from the slopes of these plots. For each bolometer timeline a temperature component is obtained from the multiplication of $\alpha_i$ with the thermistor. This temperature component is then simply subtracted from the bolometer timeline to yield the temperature-drift corrected bolometer timeline. To avoid introducing excess noise, each thermistor is low-pass filtered at 20\,mHz first. 
Since gradients are not well represented in Fourier space, a 5$^{\rm th}$ order polynomial is fit to the thermistor signals and subtracted before filtering, and then added back afterwards. 

The corrected timelines are shown in the bottom panels of Fig.~\ref{fig:corr}. The 250\,$\mu$m and 350\,$\mu$m flat timelines indicate that most of the correlated signal has been removed.
However, this is less true at 500\,$\mu$m. The temperature-corrected dark 500\,$\mu$m bolometer (not shown in the plot) 
has a residual similar to the one at 250 and 350\,$\mu$m, suggesting that the excess signal in the light detectors is optical pick-up. It is possible that the 500\,$\mu$m array is sensitive to the  blackbody light radiated by the $\la 4.5$\,K 
optics box cavity, which is not traced by the thermistor. At 500\,$\umu$m this is 1.5 per cent of the peak emission (in $\nu I_\nu$ of a 4.5\,K blackbody), while it is negligible at the shorter wavelengths. Although it might be possible to decorrelate a second component from the 500\,$\mu$m timelines, this turns out to not be necessary, as shown in the following Section. We also notice that this residual is present only where the temperature monitored by the thermistor has a gradient in excess of $\sim 1$\,nV\,s$^{-1}$, measured in the timeline units. 

It should be noted that dark bolometers might also be used to decorrelate the temperature component from the timelines with similar results. This is useful when the thermometer timelines are not available. However we have not addressed in our study the possible issue of light cross-talk between light and dark bolometers.

\section{Noise analysis}\label{sec:noise_analysis}
\begin{figure}
 \includegraphics[width=\linewidth]{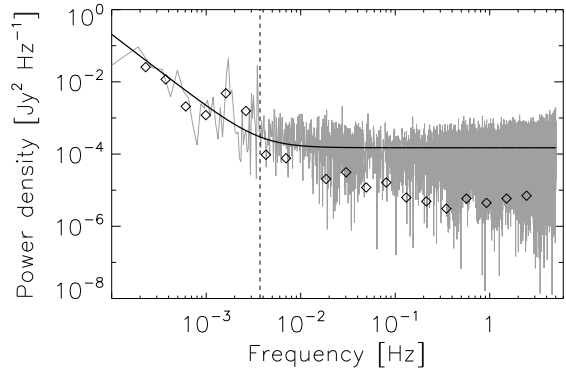}
 \caption{The noise power spectral density of a 
representative 250\,$\mu$m detector is shown in the top panel.  The solid line 
is a fit of Brown and white noise to the data, showing no evidence of steeper power-law components. 
The noise 'knee' is indicated by the vertical dashed line and is in general below
$4$\,mHz for the detectors in the three SPIRE arrays, which indicates 
a photometer temporal stability  $\ga 250$\,s. The feature at 1.7\,mHz (and its harmonics) is at the telescope scan frequency. The logarithmically binned noise cross-power 
spectrum  between two 250\,$\mu$m detectors (diamond symbols) shows that correlations among detectors becomes negligible above the noise 'knee'.}
\label{fig:psd}
\end{figure}
To understand the noise properties of the clean, temperature-corrected bolometers, it is necessary to estimate noise timelines. This is done by suppressing the astronomical signal in the timelines, because cirrus and point sources  have a contribution which can be above the noise in individual samples. 
Naive maps with 6, 10, and 14 arcsec pixels, at 250, 350, and 500\,$\mu$m, respectively, are generated by binning the timelines of the two SDP observations. These pixel sizes are the default in the standard pipeline. New timelines are obtained for each detector by 
re-scanning the map with the telescope pointing solution, effectively simulating new observations. Since each map pixel contains $\sim 10$ samples, the instrumental noise in the re-scanned timelines is about 1/3 that of the original timelines.  The re-scanned timelines are subtracted from the cleaned bolometers to provide estimates of detector noise. The noise power spectral density (PSD) is shown in Fig.~\ref{fig:psd}, for a representative 250\,$\mu$m detector.

A simple $1/f$ noise model is fit to the PSD
\begin{equation}
 P(f) = w_0^2\;\;\left[\left(\frac{f_0}{f}\right)^2 + 1\right] 
\end{equation}
where $w_0$ is the white noise level, and $f_0$ is the 'knee' of the $1/f$ noise.

The model fit values for $f_0$ are below 4\,mHz across the arrays, corresponding
to angular modes larger than $4.2^\circ$ (i.e. the same as the survey angular size), at the scan rate of 60 arcsec\,s$^{-1}$. The estimates of the white noise level are, $w_0 \simeq $ 8.6, 9.3, and 13.2\,mJy\,s$^{0.5}$ (10 percent estimated variability across the arrays) at 250, 350 and 500\,$\mu$m, which are compatible with the results of \citet{nguyen2010}, within their quoted uncertainties.

Residual correlations among detectors are investigated by estimating the cross-power spectum between channels in an array. An example for two representative 250\,$\mu$m detectors is also shown in Fig.~\ref{fig:psd}. It is significant at frequencies below the noise 'knee', meaning that the residual $1/f$ noise is highly correlated among detectors. This is not a problem for map making because these temporal scales (indicated by a vertical dotted line in the Figure) can be safely filtered, as they all correspond to unconstrained angular modes (next Section). However, we notice that the cross-correlation is significantly different from zero at high frequency, indicating the presence of broad-band correlated noise of instrumental origin. The same level of correlation is observed in the raw timelines, indicating that this is not an artifact introduced by our analysis.
The level of this correlation is estimated at 1\,Hz as the ratio of the cross-power spectrum to the PSD, which is 2, 10, and 25 per cent, at 250, 350 and 500\,$\mu$m, respectively, and negligible among detectors in different arrays. The origin of the effect is unclear, however it is not a concern for map making because the correlated noise component is sub-dominant relative to the uncorrelated component, and because different detectors in an array point at different directions in the sky, at any given time. 

\section{Maps}
\begin{figure}
 \includegraphics[width=\linewidth]{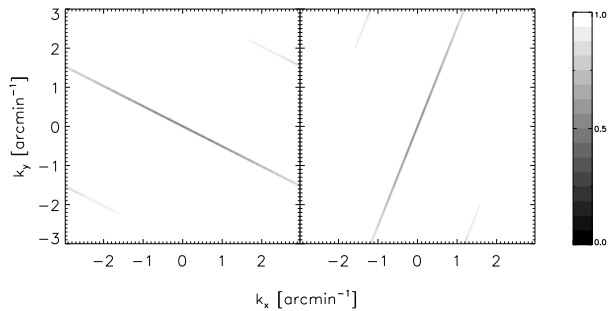}
\caption{The map maker transfer functions in Fourier space are plotted for
the two SDP scan maps in the left and right panels. Fourier modes are
attenuated and suppressed in the cross scan directions only, where the 
transfer function is close to zero. Fourier modes along the telescope scan
directions are not attenuated. Combining the two scans allows for optimal
reconstruction of all the accessible Fourier modes.}
\label{fig:tf}
\end{figure}

A 5$^{\rm th}$ order polynomial is fit and subtracted from each temperature corrected timeline, to remove residual gradients. A line is fit to the first and last 
10\,sec of data of each timeline, and subtracted.
Temporal frequencies longer than the $1/f$  'knee' are suppressed by high-pass filtering each timeline at the frequency of 4\,mHz, which corresponds to 4.2$^\circ$ on the sky, at the survey scan rate. This preserves all the accessible angular modes in one observation (the modes along the scan direction) while suppressing the correlated noise which would otherwise complicate the analysis or introduce artifacts. The weak filtering adopted does not distort the PSF in a significant way, as discussed in the next Section. This set of timelines is the final, improved Level-1 data product, which forms the input to the map maker.

The naive map maker implemented in the SPIRE standard pipline is used to bin the improved Level-1 data into maps with pixel sizes of 5, 10, and 10 arcsec, at 250, 350, and 500\,$\mu$m, with identical astrometry.
This pixelization makes it trivial to rebin the 250\,$\mu$m map onto the 350 and 500\,$\mu$m grid, when needed, without loss of information, and Nyquist sample the PSF at each wavelength.
These pixel sizes are smaller than that used by the standard pipeline (6, 10 and 14 arcsec), and caution should be taken as a few pixels will have no bolometer samples (i.e. zero coverage), which means they will be left blank.
This is not a problem because blank pixels are properly flagged in a mask which is also generated by the map maker.
Three maps are generated: the first two are obtained by binning data from the two individual, cross-linked observations (M$_1$ and M$_2$), and the other is a combined map (M$_{\rm c}$) made using timelines from both observations. Coverage and noise maps are also generated.

The two separate SDP scan maps are similar in coverage, and number of samples in each pixel. Therefore, it is reasonable to assume similar noise properties in both. By subtracting one map from the other, we obtain a jackknife which does not contain astronomical signal, and can be used to characterise the instrumental component of the noise in the maps. A $\chi^2$ test of Gaussianity  on the jackknife shows that the instrumental noise is well described by a Gaussian distribution, with reduced $\chi^2 = $ 1.3, 1.4, and 1.3, and $Q(\chi^2/\nu) = $ 15, 7, and 14 per cent, for 250, 350 and 500\,$\mu$m, respectively ($\nu$ are the degrees of freedom).

The instrumental noise RMS, $\sigma$, is estimated from the jackknife pixel histogram, and we find $\sigma = $ 29.6, 31, and 36\,mJy\,$\sqrt{\rm sample}$ (i.e. the standard deviation of each sample in a timeline), at 250, 350 and 500\,$\mu$m, respectively. These values are compatible with the noise estimated from the timelines, within $\sim$15 per cent, and are in good agreement with  the noise estimated by the map maker, which is provided in the form of a noise map (estimated from the variance in a pixel, divided by the number of samples in the pixel).

We also estimate the confusion noise, the signal arising from faint extragalactic sources blended together by the finite telescope PSF. Smoothed maps are produced by the inverse variance weighted convolution with the telescope PSF,
and a region of low cirrus emission is selected, from which a pixel histogram is formed. This histogram represents the  combined contributions of instrumental noise, confusion noise, and emission from detected sources within the region \citep{patanchon2009, marsden2009}. A Gaussian function is fit to the negative part of the histogram only, whose width is the square root of the quadrature sum of confusion and instrumental noise, the latter estimated from the pixel histogram of the smoothed jackknife. 

The 1-$\sigma$ instrumental noise in the PSF-convolved map is estimated to be 4.8, 4.9, and 5.7\,mJy\,beam$^{-1}$ at 250, 350 and 500\,$\mu$m, respectively, with 5 per cent uncertainty. The confusion noise is, respectively, 5.3, 6.4, and 6.7\,mJy\,beam$^{-1}$, with a 7 per cent uncertainty (calibration uncertainties are not included). These can be considered as the relevant values when extracting point sources from the maps.

\section{Calibration}\label{sec:calibration}
The astrometry has been verified using the point sources detected in the 250\,$\mu$m SDP maps \citep{rigby2010}. We produced histograms of the separations in RA and Dec for all SDSS DR7 r-band sources within 50 arcsec of the 250\,$\mu$m centroids, in bins of 1 arcsec. We then fit the resulting distribution of separations using a Gaussian model for the SPIRE positional errors and allowing for the effects of galaxy clustering in the SDSS data \citep[see][]{smith2010}. We find a small, $\la 2$ arcsec astrometric shift ( $\la 1$ arcsec uncertainty), which is compatible with pixelization effects, and is accounted for by shifting the maps. A similar analysis using stacking of a catalogue of IRAS sources, with positions determined in the optical and radio, gives compatible results at each wavelength. It is interesting to note that \cite{pilbratt2010} report a 1-$\sigma$ pointing uncertainty of $\sim 2$ arcsec, similar to what we find, although stochastic in nature.

The flux calibration is carried out in the timelines as part of the standard SPIRE data reduction (as discussed in Section \ref{sec:datapreprocessing}), and the resulting maps are calibrated in Jy/beam. 
We applied the flux correction factors of \cite{swinyard2010} and \cite{griffin2010} required by the SPIRE calibration available at the time of this work. The 250, 350 and 500\,$\mu$m maps are multiplied by 1.02, 1.05, and 0.94, respectively\footnote{ See http://herschel.esac.esa.int/SDP$\_$wkshops/presentations/IR/\\3$\_$Griffin$\_$SPIRE$\_$SDP2009.pdf.}.
The map making process effectively convolves the sky emission with the pixel window function, introducing a gain variation for point sources of up to 5 per cent for the 250\,$\mu$m array. Instead of accounting for this effect in the maps, we prefer to include it in the PSFs used to retrieve point source fluxes. Gaussian PSFs are fit to maps of Neptune, a SPIRE primary calibrator. The PSFs are then normalized using the Neptune flux model of \cite{moreno2010}, calculated for the day when Neptune was observed.
The PSFs, available as part of the SDP release, have FWHM of 18.1, 24.8 and 35.2 arcsec, and peak amplitudes of 0.95, 0.95 and 1.02 at 250, 350 and 500\,$\mu$m, respectively. Future release of H-ATLAS products will make use of PSFs estimated from Neptune maps.

\section{Sky reconstruction}\label{sec:sky_reconstruction}
\begin{figure}
 \includegraphics[width=\linewidth]{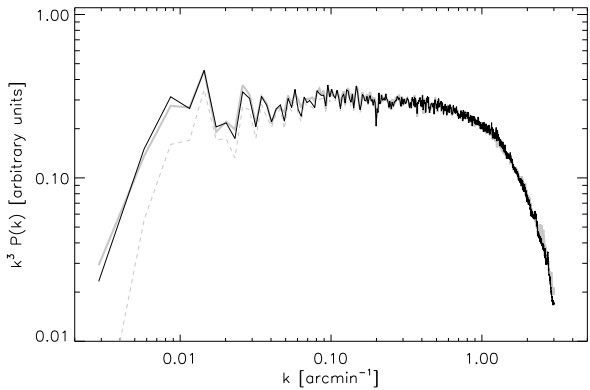}
\caption{The 1-D power spectrum of a simulated
map of Gaussian noise with  
a $k^{-3}$ power spectrum is shown by the grey solid line. The reconstructed map power spectrum
(dashed line) shows the effect of the timeline filtering in suppressing
the larger angular scales. Deconvolving this attenuation by using
the map transfer functions for the two orthogonal scans recovers all the 
accessible angular scales, as shown by the solid black line.
The drop off at small scales (large $k$) is caused by the PSF.}
\label{fig:mapps}
\end{figure}
In this section we investigate how well the pipeline is able to provide an estimate of the sky emission.
As the noise has already been discussed, it is now necessary to study the effect that high-pass filtering the timelines has on the maps. This operation is equivalent to convolving the sky with an effective filter, or transfer function:
\begin{equation}
 {\rm M}_n = T_n\otimes S + {\rm noise}, 
\end{equation}
where $n$ is an index identifying the map (1: first observation, 2: second observation, and c: combined), $T_n$ is the effective transfer function, and  $\otimes$ is the convolution operator.

As mentioned in Section~\ref{sec:mapmaking}, Fourier modes orthogonal to the scan direction are lost in the $1/f$ noise, which is filtered in the timelines. The two transfer functions for each observation are shown in Fig.~\ref{fig:tf}. These have been estimated from 900 realizations of white noise sky emission, reconstructed by the pipeline, and show how the largest Fourier angular modes are the most affected.

The maps preserve a unit response to point sources. To verify this, noiseless maps with Gaussian PSFs of nominal sizes for the three arrays are generated. The timelines obtained by re-scanning the maps are then processed by the pipeline into new maps. From the comparison between the input and output maps, we find that the gain variation is below 1 per cent and therefore negligible, compared to other sources of error, particularly the 15 per cent calibration uncertainty estimated by \citet{swinyard2010}. 

The large angular scale Fourier modes which are lost in one observation, are constrained by the other. It is therefore possible to combine the two maps with no loss of information, 
when the transfer functions are known.  A Maximum Likelihood (ML) estimator of the sky is in this case (assuming a constant and equal noise in each map)
\begin{equation}\label{eq:mapml}
 {\tilde{S}} = \frac{{\rm M}_1\; {T}_1^* +{\rm M}_2\; {T}^*_2}{T_1\,T_1^* + T_2\,T_2^*},
\end{equation}
where the symbols in the equation are the corresponding Fourier transforms of the quantities introduced earlier, and the complex conjugate is denoted by an asterisk.

To verify how effective this estimator is, tests were performed using a noiseless sky realization of Galactic cirrus emission, assuming a $k^{-3}$ power spectrum \citep[e.g.][]{miville2007}, where $k$ is the inverse angular scale. After reconstructing the emission with our pipeline, the input and reconstructed power spectra were compared  (Fig.~\ref{fig:mapps}). The effect of attenuating the large scales is shown by the dashed line. This affects Fourier modes which are larger than 20 arcmin on the sky (and therefore irrelevant for point sources). 
However, the power spectum of the ML estimator in  Equation~\ref{eq:mapml} (shown by the black line in the figure) demonstrates that all accessible Fourier modes have been properly reconstructed. 

It should be noted that the scientific analysis of H-ATLAS data taken during the SDP phase were constrained to point sources, or moderately $<2$ arcmin extended sources only. Since these sources  are not affected by this choice of filtering, the naive maps were used, instead of the ML ones described here. 

Because detectors can be effectively de-correlated, it would be possible to use the well-tested Cosmic Microwave Background code MADmap \citep{cantalupo2010} in the future. This is an optimal map maker, which does not currently handle detector-detector correlations, and does not require an estimate of the map making transfer function to compute the GLS sky estimator. The detector noise can be made stationary (at least over the time spanned by our data-set) and the noise model required by MADmap can be efficiently pre-computed and consistently used to reduce the whole H-ATLAS $\sim 550$\,deg$^2$ data-set, when it becomes available. A MADmap implementation  is now being optimised in the SPIRE pipeline, and preliminary tests suggest that it takes $\la 30$\,min to reduce the SDP dataset on a single CPU machine, provided that enough memory is available for the data.

\section{Conclusions}
Data from the SDP H-ATLAS observations have been reduced into maps using a customised pipeline. Noise analysis of individual bolometer timelines demonstrates that the SPIRE photometer is very stable, with $1/f$ residual noise dominating frequencies below $4$\,mHz. At the scan velocity adopted for this survey, this corresponds to the angular scale of the survey region. Naive map making can therefore be used to provide the optimal reconstruction of point source sky emission. However, to constrain the large scale structure it has been necessary to estimate the map transfer function by using computationally-intensive Monte Carlo techniques. 

The H-ATLAS maps contain Gaussian instrumental noise, in addition to extragalactic confusion noise, and the two have roughly the same contribution. Estimates of the confusion noise are in agreement with those measured from alternative studies. Residual correlations among detectors in the same array are found to be sub-dominant compared to the uncorrelated (among detectors) noise component, at all relevant time scales. 

This map making technique represents an intermediate solution until the implementation of MADmap in the SPIRE standard pipeline is completely characterised and tested. 
Nevertheless, the improved Level-1 data products, and noise models produced in this work will provide the  necessary inputs in order for MADmap to perform optimally.
This seams likely to be the method-of-choice to reduce the $\sim 550$\,deg$^2$ H-ATLAS survey, as it does not require the explicit estimation of the transfer functions, and detectors can be efficiently de-correlated at the time and angular scales of interest, and moreover the noisy, correlated long temporal scales can be filtered without introducing artifacts. 

\section*{Acknowledgements}
The {\it Herschel}-ATLAS is a project with {\it Herschel}, which is an ESA space observatory with science instruments provided by Europeanled Principal Investigator consortia and with important participation from NASA. U.S. participants in {\it Herschel}-ATLAS acknowledge NASA support through a
contract from JPL. We would like to thank George Bendo, Matt Griffin and Andreas Papageorgiou for the helpful discussions.

\label{lastpage}


\begin{thebibliography}{99}
\bibitem[\protect\citeauthoryear{Bock et al.}{1996}]{bock1996} Bock J.~J., Del Castillo H.~M., Turner A.~D., Beeman J.~W., Lange A.~E., Mauskopf P.~D., 1996, ESASP, 388, 119 
\bibitem[\protect\citeauthoryear{Bock et al.}{1998}]{bock1998} Bock J.~J., Glenn J., Grannan S.~M., Irwin K.~D., Lange A.~E., Leduc H.~G., Turner A.~D., 1998, SPIE, 3357, 297 
\bibitem[\protect\citeauthoryear{Cantalupo et al.}{2010}]{cantalupo2010} Cantalupo C.~M., Borrill J.~D., Jaffe A.~H., Kisner T.~S., Stompor R., 2010, ApJS, 187, 212
\bibitem[\protect\citeauthoryear{Chapin et al.}{2010}]{chapin2010}  Chapin E.~L., et al., 2010, arXiv, arXiv:1003.2647 
\bibitem[\protect\citeauthoryear{Devlin et al.}{2009}]{devlin2009} Devlin M.~J., et al., 2009, Natur, 458, 737 
\bibitem[\protect\citeauthoryear{Driver et al.}{2009}]{driver2009} Driver S.~P., et al., 2009, IAUS, 254, 469 
\bibitem[\protect\citeauthoryear{Eales et al.}{2010}]{eales2010} Eales S., et al., 2010, PASP, 122, 499 
\bibitem[\protect\citeauthoryear{Griffin et al.}{2010}]{griffin2010} Griffin M.~J., et al., 2010, arXiv,  arXiv:1005.5123 
\bibitem[\protect\citeauthoryear{Griffin et al.}{2008}]{griffin2008} Griffin M., et al., 2008, SPIE, 7010
\bibitem[\protect\citeauthoryear{Ibar et al.}{2010}]{ibar2010} Ibar E., et al., 2010, arXiv, arXiv:1009.0262 
\bibitem[\protect\citeauthoryear{Lupton}{1993}]{lupton1993} Lupton R. 1993, Statistics in theory and practice (Princeton: Princeton University Press)
\bibitem[\protect\citeauthoryear{Marsden et al.}{2009}]{marsden2009} Marsden G., et al., 2009, ApJ, 707, 1729 
\bibitem[\protect\citeauthoryear{Miville-Desch{\^e}nes et al.}{2007}]{miville2007} Miville-Desch{\^e}nes M.-A., Lagache G., Boulanger F., Puget J.-L., 2007, A\&A, 469, 595 
\bibitem[\protect\citeauthoryear{Moreno}{2010}]{moreno2010} Moreno R., 2010, Neptune and Uranus planetary brightness temperature tabulation, available from ESA Herschel Science Centre.
\bibitem[\protect\citeauthoryear{Nguyen et al.}{2010}]{nguyen2010} Nguyen H.~T., et al., 2010, arXiv, arXiv:1005.2207 
\bibitem[\protect\citeauthoryear{Pascale et al.}{2009}]{pascale2008} Pascale E., et al., 2009, ApJ, 707, 1740 
\bibitem[\protect\citeauthoryear{Patanchon et al.}{2008}]{patanchon2008} Patanchon G., et al., 2008, ApJ, 681, 708 
\bibitem[\protect\citeauthoryear{Patanchon et al.}{2009}]{patanchon2009} Patanchon G., et al., 2009, ApJ, 707, 1750 
\bibitem[\protect\citeauthoryear{Pascale et al.}{2009}]{pascale2009} Pascale E., et al., 2009, ApJ, 707, 1740 
\bibitem[\protect\citeauthoryear{Pilbratt et al.}{2010}]{pilbratt2010} Pilbratt G.~L., et al., 2010, arXiv, arXiv:1005.5331 
\bibitem[\protect\citeauthoryear{Poglitsch et al.}{2010}]{poglitsch2010} Poglitsch A., et al., 2010, arXiv, arXiv:1005.1487
\bibitem[\protect\citeauthoryear{Rigby et al.}{2010}]{rigby2010} Rigby E., et al., 2010, MNRAS, submitted 
\bibitem[\protect\citeauthoryear{Rownd et al.}{2003}]{rownd2003} Rownd B., Bock J.~J., Chattopadhyay G., Glenn J., Griffin M.~J., 2003, SPIE, 4855, 510 
\bibitem[\protect\citeauthoryear{Smith et al.}{2010}]{smith2010} Smith D.~J.~B., et al., 2010, arXiv, arXiv:1007.5260 
\bibitem[\protect\citeauthoryear{Swinyard et al.}{2010}]{swinyard2010} Swinyard B.~M., et al., 2010, arXiv, arXiv:1005.5073 



\end{thebibliography}
\end{document}